\documentclass[12pt]{article}
\usepackage{amsmath,latexsym,amssymb,psfrag,cite}

\newcommand{\be}{\begin{equation}}  
\newcommand{\ee}{\end{equation}}  
\newcommand{\bea}{\begin{eqnarray}}  
\newcommand{\eea}{\end{eqnarray}}  
\renewcommand{\Re}{\operatorname{Re}}
\renewcommand{\Im}{\operatorname{Im}}

\begin{document}

\thispagestyle{empty}
\vspace*{.5cm}
\noindent
HD-THEP-05-15 \hfill 14 July 2005

\vspace*{1.9cm}

\begin{center}
{\Large\bf K\"ahler Corrections\\[.5cm] for the 
Volume Modulus of Flux Compactifications}
\\[2.3cm]
{\large G.~von Gersdorff$\,^a$ and A.~Hebecker$\,^b$}\\[.5cm]
{\it $^a$Department of Physics and Astronomy, Johns Hopkins University, 
3400 N Charles Street, Baltimore, MD 21218}\\[.2cm]
{\it $^b$Institut f\"ur Theoretische Physik, Universit\"at Heidelberg,
Philosophenweg 16 und 19, D-69120 Heidelberg, Germany}
\\[.4cm]
{\small\tt (\,gero@pha.jhu.edu\,,\,
a.hebecker@thphys.uni-heidelberg.de\,)}
\\[1.3cm]

{\bf Abstract}
\end{center} 
No-scale models arise in many compactifications 
of string theory and supergravity, the most prominent recent example being 
type IIB flux compactifications. Focussing on the case where the no-scale 
field is a single unstabilized volume modulus (radion), we analyse the 
general form of supergravity loop corrections that affect the no-scale 
structure of the K\"ahler potential. These corrections contribute to the 4d 
scalar potential of the radion in a way that is similar to the Casimir effect. 
We discuss the interplay of this loop effect with string-theoretic $\alpha'$ 
corrections and its possible role in the stabilization of the radion.

\newpage
In flux compactifications of type IIB supergravity, all complex
structure moduli and the dilaton are generically fixed by the
non-trivial superpotential induced by the 3-form field
strength~\cite{Polchinski:1995sm, Giddings:2001yu}. However, this
superpotential is independent of the K\"ahler moduli. Even if
supersymmetry is broken by the non-zero vacuum expectation value of
the superpotential $W$, one of the flat directions associated with the
K\"ahler moduli survives. The resulting 4d model is of no-scale type
and the no-scale field $T$ is the K\"ahler modulus related to an
overall rescaling of the compact volume.  Perturbative corrections
generically renormalize the K\"ahler potential, destroy the no-scale
structure and lift the flat directions.  We will be interested in loop
corrections to the no-scale K\"ahler potential of the volume modulus
$T$ (the radion).  In the large-volume limit, such corrections should
be calculable within the low-energy effective field theory. They are
potentially relevant for the stabilization of the radion and the
uplifting to a metastable de Sitter
vacuum~\cite{Kachru:2003aw,Burgess:2003ic}.

To understand the supergravity 1-loop corrections, we first focus on a 
situation where $W=0$ and supersymmetry is unbroken. We consider the 
corrections to the radion kinetic term and to the Einstein-Hilbert term of the 
4d effective theory. Before Weyl rescaling, these corrections are independent 
of the 10d Planck mass and their form can therefore be inferred from 
dimensional arguments. It is then straightforward to derive the 
corresponding K\"ahler corrections, which are of the form $1/(T+\bar{T})^2$,
with the compact volume scaling as $V\sim ($Re$\,T)^{3/2}$. After a small 
non-zero $W$ has been introduced as a perturbation, they induce a potentially 
important contribution to the radion scalar potential.

For the purpose of our technical discussion, we first adopt a slightly more 
general perspective. Consider a $d$-dimensional supergravity theory, 
compactified to 4d on a $k$-dimensional manifold ($d=4+k$) for which its 
total volume $V$ corresponds to a flat direction. We write the metric as
\be
d s^2=g_{\mu\nu}dx^\mu dx^\nu+R(x)^2\tilde g_{mn}dy^mdy^n\,,
\ee
where Greek and Latin indices run over $0...3$ and $5...d$ 
respectively and the decomposition $g_{mn}=R^2\tilde{g}_{mn}$ is defined in 
such a way that the volume of the compact space measured with the metric 
$\tilde{g}_{mn}$ is 1. The physical volume is $V=R^k$. In spite of its 
various interesting physical effects~\cite{Giddings:2001yu}, we neglect for 
simplicity the possible warp factor, i.e., we assume that $g_{\mu\nu}$ 
does not depend on $y$. This is justified in the large volume limit. 

Assuming that the fundamental $d$-dimensional Einstein-Hilbert term has 
coefficient $M^{d-2}/2$, the 4d action reads
\be
S=\int d^4x\sqrt{g} (MR)^k \frac{M^2}{2}\left[{\cal R}+k(k-1)
(\partial\ln R)^2+\cdots\right]\,,\label{bw}
\ee
where ${\cal R}$ is the 4d curvature scalar. A possible dilaton dependence 
of the coefficient $M^{d-2}$ has not been made manifest since we assume 
that the dilaton (as well as other moduli) are stabilized at a high scale. 

We now set $M=1$ and perform a Weyl rescaling 
\be
g_{\mu\nu}\to R^{-k}g_{\mu\nu}\,,
\label{weyl}
\ee
which takes us to the Einstein frame action 
\be
S=\int d^4x\sqrt{g}\left[\frac{1}{2}{\cal R}-\frac{k(k+2)}{4}
(\partial \ln R)^2+\cdots\right]\,.\label{aw}
\ee
Note that the reason for our very explicit derivation of this familiar 
action is the importance of the intermediate form, Eq.~(\ref{bw}), for the 
subsequent discussion of quantum corrections. 

We further assume that the effective 4d theory is an $N=1$ no-scale
model~\cite{ns} where the flat direction $R$ is described by a no-scale field
$T$ with $\Re T=R^\alpha$. By comparing Eq.~(\ref{aw}) with the
kinetic term derived from the standard no-scale K\"ahler potential
$K=-3\ln(T+\bar T)$, we find
\be
\alpha=\sqrt{k(k+2)/3}\,.
\label{alpha}
\ee
Although our analysis is general, we will primarily focus on two cases: 
\begin{itemize}
\item
$d=5$ $(k=1)$ compactifications of minimal 5d supergravity on
$S^1/Z_2$ with supersymmetry broken by the Scherk-Schwarz
mechanism~\cite{Scherk:1978ta}. In this case, the no scale field is
given by $T=R+i A_5$, $A_5$ being the fifth component of the
graviphoton. This is in agreement with the value $\alpha=1$ implied by
Eq.~(\ref{alpha}). The constant superpotential characteristic of the
no-scale model is proportional to the Scherk-Schwarz parameter. This
simple and familiar example will provide us with a useful consistency
check for our results.\footnote{For the relation of 5d Scherk-Schwarz breaking
to 4d no-scale models see Refs.~\cite{Luty:1999cz}.}
\item
$d=10$ $(k=6)$ flux compactifications of type IIB string theory, where
the internal compact space is a Calabi-Yau orientifold. These indeed
result in a no-scale model with no-scale-field $T=R^4+
ib$~\cite{Giddings:2001yu}, where $b=\Im T$ stems from the dimensional
reduction of the RR four form. Note that, again, the relation between
$\Re T$ and $R$ is correctly given by the exponent of
Eq.~(\ref{alpha}).
\end{itemize}
Perturbative corrections $\Delta K$ to the K\"ahler potential generically 
destroy the no-scale structure. After supersymmetry is broken by the 
addition of a constant superpotential $W$ (which we consider to be a 
parametrically small effect), this K\"ahler correction generates a 
nontrivial potential for the volume modulus. A common approach in 
field-theoretic model building is to calculate this potential (i.e., the 
Casimir energy) and, if required, to infer the corresponding K\"ahler
correction (see, e.g.,~\cite{casimir} and, in particular,~\cite{Luty:2002hj}). 

Here, we instead consider the K\"ahler correction directly in the model 
with $W=0$, i.e., before SUSY breaking. We identify the structure 
of $\Delta K$ from the corrections to the kinetic term of the field $R$
and to the 4d Einstein term. These corrections are most easily understood 
in the 4d action before Weyl rescaling, Eq.~(\ref{bw}). 

The tree-level action is invariant under shifts in $\Im T$ and we can
expect the 1-loop correction to respect this symmetry. The finite
corrections, to be added to the action of Eq.~(\ref{bw}), then take
the form
\be 
\Delta S = \int d^4x\sqrt{g}\left[F(R){\cal R}+G(R)(\partial R)^2\right]
\,,\label{kin}
\ee
i.e., there is no explicit dependence on $\Im(T)$. 

The form of the functions $F$ and $G$ follows from dimensional arguments.
At 1-loop, the corrections arise simply from the propagation of 
$d$-dimensional free fields in the compact space. Alternatively, one may 
say that they arise from a summation of a Kaluza-Klein tower of 4d fields
with mass splitting $\sim 1/R$. In any case, the $d$-dimensional Planck 
mass $M$ does not enter these corrections and the only scale known to these 
corrections is the compactification radius $R$. Thus, from the requirement 
of a dimensionless 4d action, we have $F(R)\sim 1/R^2$ and $G(R)\sim 1/R^4$, 
i.e., the corrections are 
\be
R^{-2}{\cal R} \qquad\mbox{and}\qquad  R^{-4}(\partial R)^2\,.
\label{operators}
\ee

In the above, we have pretended that the field-theoretic one-loop
corrections are finite. If they are not, a UV cutoff scale (say the
string scale $\alpha'$) enters the result. However, such
cutoff-dependent contributions can always be absorbed in a local
$d$-dimensional action (including higher-dimension operators). The
leading operators relevant for us are those of Eq.~(\ref{bw}) (a
$d$-dimensional cosmological constant is not generated in
supersymmetric theories).  Subdominant terms may be important. For
example, the $\alpha'$ corrections of~\cite{louis} (see
also~\cite{antoniadis}) considered recently in this context
\cite{Balasubramanian:2004uy} are of this type. Our present result of
Eq.~(\ref{operators}) is limited to those corrections which cannot be
viewed as the dimensional reductions of $d$-dimensional local
operators.

The terms in Eq.~(\ref{operators}) correspond to the operators before Weyl
rescaling. Going to the Einstein frame,\footnote{Note that the Weyl
rescaling has to be modified in the presence of the first operator in
Eq.~(\ref{operators}).} we find that both operators give corrections
to the kinetic term
\be 
(R^{-(k+4)}+\dots)(\partial R)^2 \,,
\label{kin2}
\ee
where $\dots$ stand for terms which are suppressed by inverse powers
of $R$ in the limit $R\to\infty$.  Rewriting Eq.~(\ref{kin2}) in
terms of $T$, we see that we need a $\Delta K$ which induces a kinetic term
\be
(T+\bar T)^{-\frac{k+2}{\alpha}-2}\,\partial T\cdot\partial \bar T\,.
\ee
We conclude that
\be
\Delta K\sim \frac{1}{(T+\bar T)^c} \qquad\mbox{with}\qquad c=
\frac{k+2}{\alpha}=\sqrt{\frac{3(k+2)}{k}}\,.
\ee
We now have all the necessary information to calculate the form of the
one-loop potential for $R$ that arises if a non-zero $W$ is included. Using 
the standard supergravity formula for the scalar potential we find the 
Einstein-frame result
\be
V^{\rm E}_{\rm Casimir}(R)\sim |W|^2 (T+\bar{T})^{-(c+3)}\,.
\ee
The numerical prefactor, which we have suppressed in the above expression, 
includes a term $c(c-1)$. This vanishes for $c=0$ and $c=1$, i.e., in the two 
cases where $\Delta K$ preserves the no-scale structure (at least in the 
large-$R$ limit). 

Returning to the frame used in Eqs.~(\ref{bw}) and (\ref{kin}) (which we will 
refer to as the Brans-Dicke frame) by undoing the Weyl rescaling 
Eq.~(\ref{weyl}), we find
\be
V_{\rm Casimir}^{\rm BD}(R)\sim |W|^2 R^{-3\alpha +k-2}\,.
\label{resultBD}
\ee
In the example of the 5d compactification on $S^1$ or $S^1/Z_2$ with 
Scherk-Schwarz SUSY breaking, this gives the well-known 1-loop potential 
$V(R)\sim |W|^2 R^{-4}$. Since the Scherk-Schwarz parameter is dimensionless, 
this correction has to behave as one would expect in a massless non-SUSY 
field theory on dimensional grounds. Indeed, the $R^{-4}$ behaviour is the 
familiar scaling of the Casimir energy, which ensures that the 4d potential 
has mass dimension 4.

In the case of 10d flux compactifications, which is our primary interest in 
this paper, we obtain a correction 
\be
V^{\rm BD}_{\rm Casimir}(R)\sim |W|^2 R^{-8}\,.
\label{sugra}
\ee
This has to be compared with the perturbative string-theoretic ($\alpha'$) 
corrections~\cite{louis,Balasubramanian:2004uy} recently considered in this 
context, which scale as 
\be
V^{\rm BD}_{\alpha'}(R)\sim |W|^2R^{-6}\,.
\label{string}
\ee
Even though our Casimir correction is subdominant, it is clearly less so than 
nonperturbative corrections to the superpotential, which are expected to be 
exponentially suppressed at large volume.

Thus, if a (meta-)stable minimum at large volume can be found in the combined 
potential 
\be
V^{\rm BD}(R)\sim |W|^2(c_{\alpha'}R^{-6}+c_{\rm Casimir}R^{-8})\,,
\ee
one may hope that this result will survive a more detailed analysis. Naively, 
such a minimum appears at 
\be
R_{\rm min}^2=\frac{4|c_{\rm Casimir}|}{3|c_{\alpha'}|}\,,
\ee
whenever $c_{\alpha'}<0$ and $c_{\rm Casimir}>0$. However, as long as 
this value can not be made parametrically large, there is clearly no
reason to neglect higher-order and non-perturbative corrections. 

A similar situation has recently been discussed in the 5d
field-theoretic context, where the interplay of 1- and 2-loop Casimir
energy effects was used to stabilize a 5d model in a controlled
way~\cite{gersdorff} (for earlier related ideas see,
e.g.,~\cite{Albrecht:2001cp,Arkani-Hamed:1998nu}). The key there was
the possibility of finding a class of models with a hierarchy between
the coefficients of the two leading terms in the $1/R$ expansion.

The obvious parameter that could create such a hierarchy in the present 
context is the value of $\alpha'$. To see this in more detail, we write the 
type IIB supergravity action not in terms of $\alpha'$ and the dilaton, but 
rather in terms of $\alpha'$ and the 10d Planck mass $M$. Then the tree level 
part depends only on $M$ while the $\alpha'$ correction (and therefore the 
coefficient $c_{\alpha'}$) involve an explicit factor $\alpha'^3$. Since 
$c_{\rm Casimir}$ depends only on the tree-level supergravity action, we 
conclude that 
\be
(R_{\rm min}M)^2\sim 1/(M^6\alpha'^3)\,,
\ee
which will be large at small $\alpha'$. Unfortunately, this corresponds to 
the strong coupling regime of string theory. By the S self duality of the 
type IIB theory, the small-$\alpha'$ regime has a dual description with 
Regge slope $\tilde{\alpha'}\sim \alpha'^{-1}$. We expect $\tilde{\alpha'}$ 
corrections to arise in this theory, implying that the coefficient of the 
$R^{-6}$ term can never be made small. 

If an explicit calculation of $c_{\rm Casimir}$, along the lines 
of~\cite{berg}, were available in a sufficiently large class of models
(with known $c_{\alpha'}$), one could attempt to isolate geometries where the 
above minimum occurs accidentally at large $R$. Work in this direction is 
under way~\cite{koers_talks}. In the absence of a detailed study based on 
such explicit results, we can make the following proposal for how a large 
value of $R_{\rm min}$ may arise:
Recall that $c_{\alpha'}$ is proportional to the Euler number $\chi$ of the 
underlying manifold. We can now think of a topologically complicated 
space, where the two Hodge numbers $h^{1,1}$ and $h^{2,1}$ are large 
while $\chi=2(h^{1,1}-h^{2,1})$ is small. In this limit one might 
expect that, because of the large number of light fields (and the presumably 
large number of corresponding Kaluza-Klein towers), the coefficient $c_{\rm 
Casimir}$ will be large. Thus, stabilization at large $R_{\rm min}$ should 
naturally occur. 

To summarize, we have derived the parametrical form of 1-loop supergravity 
K\"ahler corrections to the volume modulus of type IIB flux compactifications. 
We have found the leading finite correction to be of the form $\Delta K \sim 
1/(T+\bar{T})^2$ with Re$\,T\sim R^4\sim V^{2/3}$. In the presence of a 
non-zero vacuum value of the superpotential $W$, this gives rise to a scalar 
potential of the form $|W|^2/R^8$, which is subdominant relative to the 
potential contribution $|W|^2/R^6$ induced by $\alpha'$ corrections. We 
note that our correction, which resembles the Casimir energy effect discussed 
extensively in field-theoretic models, is dominant for manifold with vanishing 
Euler number. Furthermore, for specific compact spaces, this Casimir 
correction may combine with the $\alpha'$ correction to ensure volume 
stabilization at large $R$. We expect the Casimir correction discussed in this 
paper to be relevant for a wide class of models and stabilization mechanisms. 

\noindent
{\bf Acknowledgements}: We would like to thank Jonathan Bagger, Boris
K\"ors, Jan Louis, Claudio Scrucca, Minho Son, Michele Trapletti and
Alexander Westphal for helpful discussions. GG is supported by the
Leon Madansky Fellowship and by NSF Grant P420-D36-2051.



\begin{thebibliography}{99}

\bibitem{Polchinski:1995sm}
  J.~Polchinski and A.~Strominger, ``New Vacua for Type II String Theory,''
  Phys.\ Lett.\ B {\bf 388} (1996) 736 [arXiv:hep-th/9510227];\\
  J.~Michelson, ``Compactifications of type IIB strings to four dimensions
   with non-trivial classical potential,'' Nucl.\ Phys.\ B {\bf 495} (1997)
  127 [arXiv:hep-th/9610151];\\
  S.~Gukov, C.~Vafa and E.~Witten,
  ``CFT's from Calabi-Yau four-folds,''
  Nucl.\ Phys.\ B {\bf 584} (2000) 69
  [Erratum-ibid.\ B {\bf 608} (2001) 477]
  [arXiv:hep-th/9906070];\\
  K.~Dasgupta, G.~Rajesh and S.~Sethi, ``M theory, orientifolds and G-flux,''
  JHEP {\bf 9908} (1999) 023 [arXiv:hep-th/9908088];\\
  S.~Gukov,
  ``Solitons, superpotentials and calibrations,''
  Nucl.\ Phys.\ B {\bf 574}, 169 (2000)
  [arXiv:hep-th/9911011];\\
  T.~R.~Taylor and C.~Vafa, ``RR flux on Calabi-Yau and partial supersymmetry 
  breaking,'' Phys.\ Lett.\ B {\bf 474} (2000) 130 [arXiv:hep-th/9912152];\\
  P.~Mayr, ``On supersymmetry breaking in string theory and its realization 
  in brane worlds,'' Nucl.\ Phys.\ B {\bf 593} (2001) 99
  [arXiv:hep-th/0003198];\\
  G.~Curio and A.~Krause,
  ``Four-flux and warped heterotic M-theory compactifications,''
  Nucl.\ Phys.\ B {\bf 602}, 172 (2001)
  [arXiv:hep-th/0012152];\\
  G.~Curio, A.~Klemm, D.~Lust and S.~Theisen, ``On the vacuum structure of 
  type II string compactifications on  Calabi-Yau spaces with H-fluxes,''
  Nucl.\ Phys.\ B {\bf 609} (2001) 3 [arXiv:hep-th/0012213];\\
  M.~Haack and J.~Louis, ``M-theory compactified on Calabi-Yau fourfolds 
  with background flux,'' Phys.\ Lett.\ B {\bf 507} (2001) 296
  [arXiv:hep-th/0103068].

\bibitem{Giddings:2001yu}
  S.~B.~Giddings, S.~Kachru and J.~Polchinski,
  ``Hierarchies from fluxes in string compactifications,''
  Phys.\ Rev.\ D {\bf 66} (2002) 106006 [arXiv:hep-th/0105097].

\bibitem{Kachru:2003aw}
  S.~Kachru, R.~Kallosh, A.~Linde and S.~P.~Trivedi, ``De Sitter vacua in 
  string theory,'' Phys.\ Rev.\ D {\bf 68} (2003) 046005
  [arXiv:hep-th/0301240].

\bibitem{Burgess:2003ic}
  C.~P.~Burgess, R.~Kallosh and F.~Quevedo, ``de Sitter string vacua from 
  supersymmetric D-terms,'' JHEP {\bf 0310} (2003) 056 [arXiv:hep-th/0309187].
\bibitem{ns}
E.~Cremmer, S.~Ferrara, C.~Kounnas and D.~V.~Nanopoulos, 
Phys.\ Lett.\ B 
{\bf 
133} (1983) 61;
J.~R.~Ellis, A.~B.~Lahanas, D.~V.~Nanopoulos and K.~Tamvakis, 
Phys.\ Lett.\ B {\bf 134} (1984) 429;
N.~Dragon, M.~G.~Schmidt and U.~Ellwanger, 
Phys.\ Lett.\ B {\bf 145} (1984) 192; 
Prog.\ Part.\ Nucl.\ Phys.\  {\bf 18} (1987) 1; A.~B.~Lahanas and D.~V.~Nanopoulos, 
Phys.\ Rept.\  {\bf 145} (1987) 1.
\bibitem{Scherk:1978ta}
  J.~Scherk and J.~H.~Schwarz,
  ``Spontaneous Breaking Of Supersymmetry Through Dimensional Reduction,''
  Phys.\ Lett.\ B {\bf 82} (1979) 60;\\
%
  J.~Scherk and J.~H.~Schwarz,
  ``How To Get Masses From Extra Dimensions,''
  Nucl.\ Phys.\ B {\bf 153} (1979) 61.

\bibitem{Luty:1999cz}
  M.~A.~Luty and R.~Sundrum,
  ``Radius stabilization and anomaly-mediated supersymmetry breaking,''
  Phys.\ Rev.\ D {\bf 62} (2000) 035008
  [arXiv:hep-th/9910202];\\
  Z.~Chacko and M.~A.~Luty,
  ``Radion mediated supersymmetry breaking,''
  JHEP {\bf 0105} (2001) 067
  [arXiv:hep-ph/0008103];\\
  D.~Marti and A.~Pomarol,
  ``Supersymmetric theories with compact extra dimensions in N = 1
  superfields,''
  Phys.\ Rev.\ D {\bf 64}, 105025 (2001)
  [arXiv:hep-th/0106256];\\
  G.~von Gersdorff and M.~Quiros,
  ``Supersymmetry breaking on orbifolds from Wilson lines,''
  Phys.\ Rev.\ D {\bf 65} (2002) 064016
  [arXiv:hep-th/0110132].

\bibitem{casimir}
  T.~Appelquist and A.~Chodos, ``The Quantum Dynamics Of Kaluza-Klein 
  Theories,'' Phys.\ Rev.\ D {\bf 28} (1983) 772;\\
  H.~Itoyama and T.~R.~Taylor, ``Supersymmetry Restoration In The Compactified
  O(16) X O(16)-Prime Heterotic String Theory,'' Phys.\ Lett.\ B {\bf 186}
  (1987) 129;\\
  I.~Antoniadis, ``A Possible New Dimension At A Few Tev,''
  Phys.\ Lett.\ B {\bf 246} (1990) 377;\\
  J.~Garriga, O.~Pujolas and T.~Tanaka, ``Radion effective potential in the 
  brane-world,'' Nucl.\ Phys.\ B {\bf 605} (2001) 192 [arXiv:hep-th/0004109];\\
  E.~Ponton and E.~Poppitz, ``Casimir energy and radius stabilization in five 
  and six dimensional orbifolds,'' JHEP {\bf 0106} (2001) 019
[arXiv:hep-ph/0105021];\\
  G.~von Gersdorff, M.~Quiros and A.~Riotto,
  ``Scherk-Schwarz supersymmetry breaking with radion stabilization,''
  Nucl.\ Phys.\ B {\bf 689} (2004) 76
  [arXiv:hep-th/0310190];\\
  E.~Dudas and M.~Quiros,
  ``Five-dimensional massive vector fields and radion stabilization,''
  arXiv:hep-th/0503157.



\bibitem{Luty:2002hj}
  M.~A.~Luty and N.~Okada, ``Almost no-scale supergravity,''
  JHEP {\bf 0304} (2003) 050 [arXiv:hep-th/0209178];\\
  R.~Rattazzi, C.~A.~Scrucca and A.~Strumia, ``Brane to brane gravity 
  mediation of supersymmetry breaking,'' Nucl.\ Phys.\ B {\bf 674} (2003) 171
  [arXiv:hep-th/0305184].

\bibitem{louis}
  K.~Becker, M.~Becker, M.~Haack and J.~Louis,
  ``Supersymmetry breaking and alpha'-corrections to flux induced
  potentials,''
  JHEP {\bf 0206} (2002) 060
  [arXiv:hep-th/0204254].

\bibitem{antoniadis}
  I.~Antoniadis, S.~Ferrara, R.~Minasian and K.~S.~Narain,
  ``R**4 couplings in M- and type II theories on Calabi-Yau spaces,''
  Nucl.\ Phys.\ B {\bf 507} (1997) 571
  [arXiv:hep-th/9707013].

\bibitem{Balasubramanian:2004uy}
  V.~Balasubramanian and P.~Berglund,
  ``Stringy corrections to Kaehler potentials, SUSY breaking, and the
  cosmological constant problem,'' JHEP {\bf 0411} (2004) 085
  [arXiv:hep-th/0408054];\\
  K.~Bobkov, ``Volume stabilization via alpha' corrections in type IIB 
  theory with fluxes,'' 
  JHEP {\bf 0505} (2005) 010 [arXiv:hep-th/0412239];\\
  V.~Balasubramanian, P.~Berglund, J.~P.~Conlon and F.~Quevedo,
  ``Systematics of moduli stabilisation in Calabi-Yau flux compactifications,''
  JHEP {\bf 0503}, 007 (2005)
  [arXiv:hep-th/0502058];\\
  J.~P.~Conlon, F.~Quevedo and K.~Suruliz,
  ``Large-volume flux compactifications: Moduli spectrum and D3/D7 soft
  supersymmetry breaking,''
  arXiv:hep-th/0505076;\\
  A.~Westphal, ``Eternal Inflation with alpha'-Corrections,''
  arXiv:hep-th/0507079.

\bibitem{gersdorff}
  G.~von Gersdorff and A.~Hebecker, ``Radius stabilization by two-loop
  Casimir energy,'', Nucl.\ Phys.\ B {\bf 720} (2005) 211
  [arXiv:hep-th/0504002].

\bibitem{Albrecht:2001cp}
  A.~Albrecht, C.~P.~Burgess, F.~Ravndal and C.~Skordis, ``Exponentially 
  large extra dimensions,'' Phys.\ Rev.\ D {\bf 65} (2002) 123506
  [arXiv:hep-th/0105261];\\
  L.~Da Rold, ``Radiative corrections in 5D and 6D expanding in winding 
  modes,'' Phys.\ Rev.\ D {\bf 69} (2004) 105015 [arXiv:hep-th/0311063].

\bibitem{Arkani-Hamed:1998nu}
  N.~Arkani-Hamed, M.~Dine and S.~P.~Martin,
   ``Dynamical supersymmetry breaking in models with a Green-Schwarz
  mechanism,''
  Phys.\ Lett.\ B {\bf 431}, 329 (1998)
  [arXiv:hep-ph/9803432];\\
  N.~Irges,
   ``Anomalous U(1), holomorphy, supersymmetry breaking and dilaton
  stabilization,''
  Phys.\ Rev.\ D {\bf 59}, 115008 (1999)
  [arXiv:hep-ph/9812338].

\bibitem{berg}
  M.~Berg, M.~Haack and B.~K\"ors,
  ``Loop corrections to volume moduli and inflation in string theory,''
  Phys.\ Rev.\ D {\bf 71} (2005) 026005
  [arXiv:hep-th/0404087].

\bibitem{koers_talks}
  B.~K\"ors, talk at ``String Phenomenology 2005'', Munich, June 2005.

\end{thebibliography}
\end{document}